\documentclass[12pt]{article}
\usepackage{amsmath,amssymb}
\usepackage{latexsym,graphicx,verbatim,enumerate}

\def\6#1{{\underline{#1}}}
\def\m6#1{{\underline{#1}\,}}

\newdimen\Tdim
\def\ispan{{\setbox0=\hbox{i}%
\Tdim\ht0\advance\Tdim\dp0\rule[-\dp0]{0pt}{\Tdim}}}
\def\jspan{{\setbox0=\hbox{j}%
\Tdim\ht0\advance\Tdim\dp0\rule[-\dp0]{0pt}{\Tdim}}}
\def\Tspan#1{{\setbox0=\hbox{#1}%
\Tdim\ht0\advance\Tdim\dp0\advance\Tdim.55ex\rule[-\dp0]{0pt}{\Tdim}\box0}}

\def\be{\begin{eqnarray}}
\def\ben{\begin{eqnarray*}}
\def\ee{\end{eqnarray}}
\def\een{\end{eqnarray*}}
\def\Tr{{\rm Tr}}

\def\p{\partial}

\def\D{\mathcal{D}}

\def\=:{=\hspace{-.7em}\raisebox{1.1ex}{.}\hspace{.1em}\raisebox{-0.2ex}{.} }

\newcommand {\beq}{\begin{eqnarray}}
\newcommand {\eeq}{\end{eqnarray}}

\makeatletter
\newcommand{\thetablename}{Table}
\def\fnum@table{\thetablename\ \thetable}
\makeatother

\begin{document}
\thispagestyle{empty}
\begin{flushright}
IFUP-TH/2009-5,~DAMTP-2009-21,~TIT/HEP-594\\
{\tt arXiv:0903.1518} \\
March, 2009 \\
\end{flushright}
\vspace{3mm}
\begin{center}
{\LARGE  
Multiple Layer Structure of Non-Abelian Vortex
} \\ 
\vspace{20mm}

{\normalsize
Minoru~Eto$^{a,b}$, 
Toshiaki~Fujimori$^c$, 
Takayuki Nagashima$^c$, \\
Muneto~Nitta$^d$,
Keisuke~Ohashi$^e$, 
and
 Norisuke~Sakai$^f$}
\footnotetext{
e-mail~addresses: \tt
minoru(at)df.unipi.it;
fujimori,
nagashi(at)th.phys.titech.ac.jp;\\
nitta(at)phys-h.keio.ac.jp;
K.Ohashi(at)damtp.cam.ac.uk;
sakai(at)lab.twcu.ac.jp.
}

\vskip 1.5em
$^a$ {\it INFN, Sezione di Pisa,
Largo Pontecorvo, 3, Ed. C, 56127 Pisa, Italy
}
\\
$^b$ {\it Department of Physics, University of Pisa
Largo Pontecorvo, 3,   Ed. C,  56127 Pisa, Italy
}
\\
$^c$ {\it Department of Physics, Tokyo Institute of
Technology, Tokyo 152-8551, Japan
}
\\
$^d$ 
{\it Department of Physics, Keio University, Hiyoshi, Yokohama,
Kanagawa 223-8521, Japan
}
\\
$^e$
{\it Department of Applied Mathematics and Theoretical Physics, \\
University of Cambridge, CB3 0WA, UK}
\\
$^f$
{\it Department of Mathematics, Tokyo Woman's Christian University, 
Tokyo 167-8585, Japan }
 \vspace{12mm}

\abstract{
Bogomol'nyi-Prasad-Sommerfield (BPS) vortices in $U(N)$ 
gauge theories have two layers corresponding to non-Abelian 
and Abelian fluxes, whose widths depend nontrivially on the 
ratio of $U(1)$ and $SU(N)$ gauge couplings. 
We find numerically and analytically that the widths differ 
significantly from the Compton lengths 
of lightest massive particles 
with the appropriate quantum number.
}

\end{center}

\vfill
\newpage
\setcounter{page}{1}
\setcounter{footnote}{0}
\renewcommand{\thefootnote}{\arabic{footnote}}


\noindent
{\it 1.\ Introduction.} 
Many important properties of Abelian (ANO) vortex 
were found \cite{de Vega:1976mi,Taubes:1979tm,Plohr:1981cy,
Perivolaropoulos:1993uj,Speight:1996px} since its discovery 
\cite{Nielsen:1973cs}. 
Recently vortices in $U(N)$ gauge theories 
(called non-Abelian vortices) were found \cite{HT,Auzzi:2003fs} 
and have attracted much attention \cite{NA-vortex}
because they play an 
important role in 
a dual picture of quark confinement \cite{Auzzi:2003fs,vm}
and are a candidate of cosmic strings \cite{cs} 
(see \cite{review} for review report).
The moduli space of $U(N)$ non-Abelian vortices was 
determined in \cite{Eto:2005yh} and study on interactions 
between non-BPS configurations started in \cite{aev}. 
Non-Abelian vortices in other gauge groups have been 
studied 
\cite{na_other_group}. 

Although there have been 
much 
progress and wide 
applications, internal structures and dependence 
on gauge coupling constants have not yet been studied 
for (color) magnetic flux tubes. 
It is particularly important to study physical widths of vortices 
qualitatively and quantitatively, although it is not easy because 
no analytic solutions are known. 
It may be tempting to speculate that the width is determined 
by the Compton lengths of lightest massive particles with 
the appropriate quantum number. 
Purpose of this letter is to clarify intricate multiple 
layer structures of non-Abelian vortices by investigating 
numerically and analytically the equations of motion. 
Non-Abelian vortices have two 
distinct widths for $SU(N)$ and $U(1)$ fluxes. 
We clarify properties of these widths by 
making use of several approximations. 
It turns 
out that non-Abelian vortices are very 
different from ANO vortices 
and have much 
richer 
internal structures.

\noindent
{\it 2.\ Vortex equations and solutions.}
Let us consider a $U(N)$ gauge theory with 
gauge fields $W_\mu$ for $SU(N)$ and $w_\mu$ for $U(1)$ 
and $N$ Higgs fields $H$ 
($N$-by-$N$ matrix) 
in the fundamental representation. 
We consider the Lagrangian ${\cal L} = K - V$ which 
can be embedded into supersymmetric theory with eight 
supercharges 
\beq
K \!&=&\! \Tr\! \left[ - \frac{1}{2g^2} (F_{\mu\nu})^2
+ \D_\mu H \D^\mu H^\dagger 
\right]
- \frac{1}{4e^2} (f_{\mu\nu})^2 \,,
\label{eq:kin_su(n)}\\
V \!&=&\!
\frac{g^2}{4} \Tr\!\left[ \left<HH^\dagger \right>^2 \right] 
+ \frac{e^2}{2}  
\left(\Tr\! \left[HH^\dagger - c {\bf 1}_N \right]\right)^2,
\label{eq:pot_su(n)}
\eeq
where $\left<X\right>$ stands for a traceless part of a 
square matrix $X$. 
Our notation is
$\D_\mu H = (\p_\mu + i W_\mu + i w_\mu {\bf 1}_N ) H$,
$F_{\mu\nu} = \p_\mu W_\nu - \p_\nu W_\mu + i \left[W_\mu,W_\nu\right]$ and
$f_{\mu\nu} = \p_\mu w_\nu - \p_\nu w_\mu$.
We have three couplings: $SU(N)$ gauge coupling $g$, 
$U(1)$ gauge coupling $e$ and Fayet-Iliopoulos parameter $c>0$.

The Higgs vacuum 
$H = \sqrt c\,{\bf 1}_N$ 
is unique and is in 
a color-flavor $SU(N)_{\rm C+F}$ locking phase. 
Mass spectrum 
is classified 
according to representations of $SU(N)_{\rm C+F}$ 
as $m_g \equiv  g \sqrt{c}$ for non-Abelian fields 
$\phi_{\rm N}=(W,\left< H\right>)$ and 
$m_e \equiv  e\sqrt {2Nc}$ for Abelian fields 
$\phi_{\rm A}=(w,{\rm Tr}(H-\sqrt{c}{\bf 1}_N))$.
The non-Hermitian part of $H$ is eaten by the $U(N)$ gauge fields.
A special case of $m_g = m_e$ \cite{HT} has been mostly considered so far, 
which is equivalently  
\beq
\gamma = 1
\quad {\rm with} \quad \gamma \equiv \frac{g}{e\sqrt{2N}} = \frac{m_g}{m_e}, 
\eeq
but we study general cases in this Letter.

Let us consider static vortex-string solutions along $x_3$-axis. 
The BPS equations
for the non-Abelian vortex are 
\beq
\bar \D H = 0,\quad
\frac{F_{12}}{m_g^2} = \frac{\left<HH^\dagger\right>}{2c},\quad 
\frac{f_{12}}{m_e^2} = \frac{\Tr\!(HH^\dagger - c{\bf 1}_N)}{2c},
\label{eq:bps}
\eeq
with $\bar \D = (\D_1 + i \D_2)/2$.
The tension of $k$-vortex is
$T_k = - c \int\! d^2x\, \Tr[f_{12}{\bf 1}_N] = 2\pi k c$.
No analytic solutions have been known whereas a numerical 
solution was first found in \cite{Auzzi:2003fs}.
For $k=1$ vortex, we take 
$H = S^{-1} H_0$, $\bar W = - i S^{-1}\bar \p S$ 
[$2 \bar W \equiv (W_1 + w_1 {\bf 1}_N) + i (W_2 + w_2 {\bf 1}_N)$] 
with diagonal matrices
$H_0 = {\rm diag}(\,r e^{i\theta},1,\cdots,1\,)$ and
$c SS^\dagger 
= e^{\left(\psi_e+\frac{1}{N}\log r^2 \right){\bf 1}_N 
+ \left(\psi_g+\frac{N-1}{N}\log r^2 \right) T}$.
Here, 
$T \equiv {\rm diag}(1,-\frac{1}{N-1},\cdots,-\frac{1}{N-1})$ 
and $\psi_e$ and $\psi_g$ are real functions of the 
radius $r >0$ of the polar coordinates $(r, \theta)$ in 
$x_1, x_2$ plane. 
Then we get
\beq
\frac{ \triangle  \psi_e}{m_e^2} 
+ \frac{1}{N}e^{-\psi_e}\left(e^{-\psi_g} + (N-1) e^{\frac{\psi_g}{N-1}}\right) = 1,
\label{eq:mm1a}\\
\frac{\triangle  \psi_g}{m_g^2}  
+ \frac{N-1}{N} e^{-\psi_e}\left( e^{-\psi_g} - e^{\frac{\psi_g}{N-1}}\right) = 0,
\label{eq:mm2a}
\eeq
with $\triangle  f(r) = \p_r(r\p_r f(r))/r$.
The boundary conditions 
are $\psi_e,\,\psi_g \rightarrow 0~(r \rightarrow \infty)$ 
and $N \psi_e, 
\frac{N}{N-1} \psi_g \rightarrow - \log r^2~(r\rightarrow 0)$.
The fluxes and the Higgs fields are expressed by
\beq
f_{12} = - \frac{1}{2}\triangle \psi_e,\quad
F_{12} = - \frac{1}{2} \triangle \psi_g T,\quad
H =\sqrt c\, {\rm diag}(h_1,h_2,\cdots,h_2) 
\eeq
with $h_1= e^{- \frac{\psi_e+\psi_g}{2}+i\theta}$ 
and $h_2 = e^{-\frac{1}{2}(\psi_e - \frac{\psi_g}{N-1})}$.
The amount of the Abelian flux is $1/N$ and 
the non-Abelian flux is $(N-1)/N$ of the ANO vortex.

We found the following theorems for Eqs.\,(\ref{eq:mm1a}) and (\ref{eq:mm2a})
\begin{enumerate}[a)]
\item $\psi_{e,g} > 0$,~~$\p_r\psi_{e,g} < 0$ and $\triangle\psi_{e,g} > 0$
\item $|h_1| < |h_2|$,~~$|h_1| < 1$ and $\p_r|h_1| > 0$
\item $\p_r|h_2| \gtreqless 0$ and $1 \gtreqless h_2 \gtreqless \sqrt{N/(N+\gamma^2-1)}$ for $\gamma \gtreqless 1$
\end{enumerate}
All these can be proved by using the following theorem for
an analytic function $f(r)$ 
satisfying $f(r) < 0\ \Rightarrow \triangle f(r) < 0$: 
If $\p_rf(0) \le 0$ and $f(\infty)=0$, then $f(r) \ge 0$ for $^\forall r \in (0,\infty)$.
In the case of $\gamma = 1$, we get $N \psi_e = \frac{N}{N-1} \psi_g
\equiv \psi_{\rm ANO}$ and
the above equations reduce to
$\triangle \psi_{\rm ANO}  
= m_e^2 (1 - e^{-\psi_{\rm ANO}})$ 
with boundary condition $\psi_{\rm ANO} \rightarrow 0~(r \rightarrow \infty)$ and $\psi_{\rm ANO} \rightarrow -\log r^2~(r \rightarrow 0)$.

\begin{figure}[ht]
\begin{center}
\begin{tabular}{lcl}
\includegraphics[width=6cm]{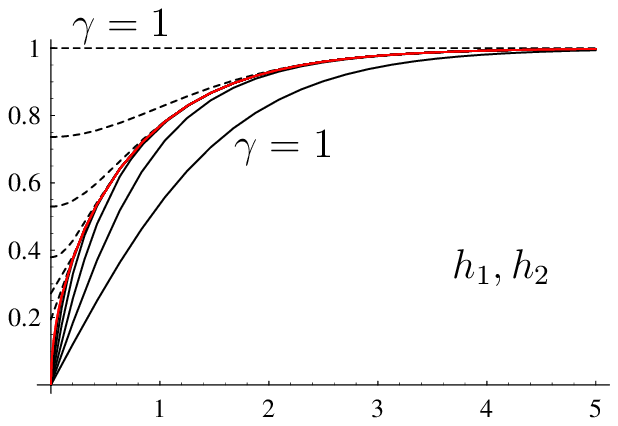} & $\ $ & \includegraphics[width=6cm]{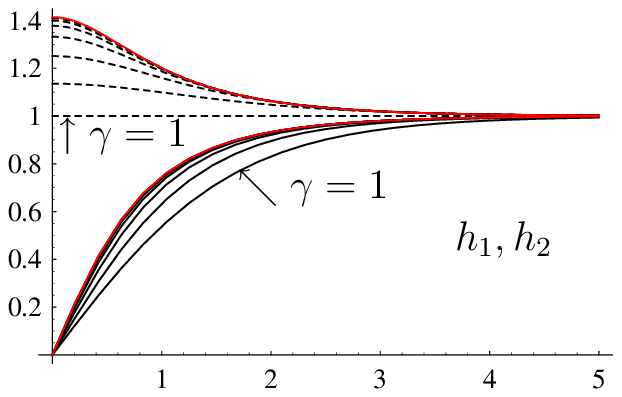}  \\
\includegraphics[width=6cm]{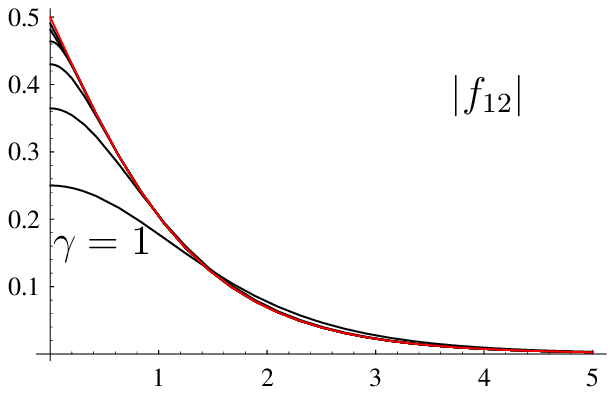} & $\ $ & \includegraphics[width=6cm]{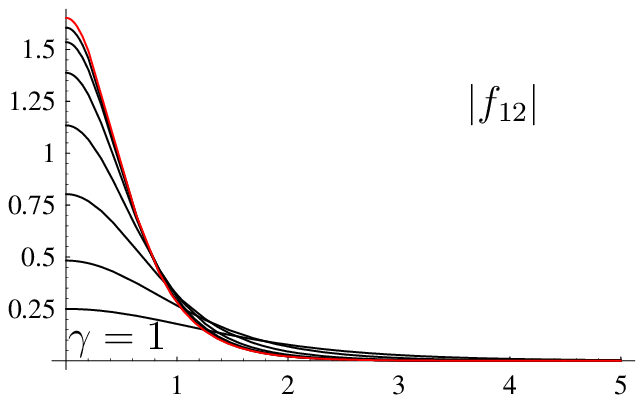}\\
\includegraphics[width=6cm]{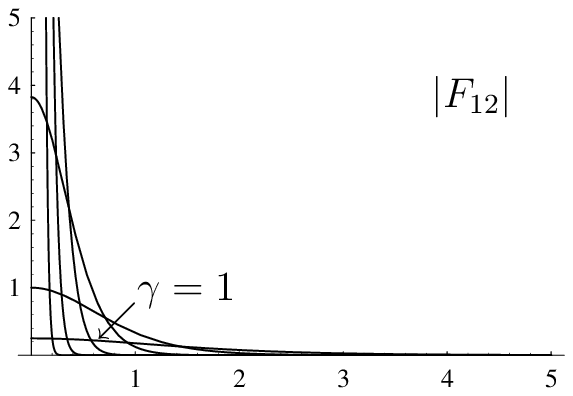} & $\ $ &  \includegraphics[width=6cm]{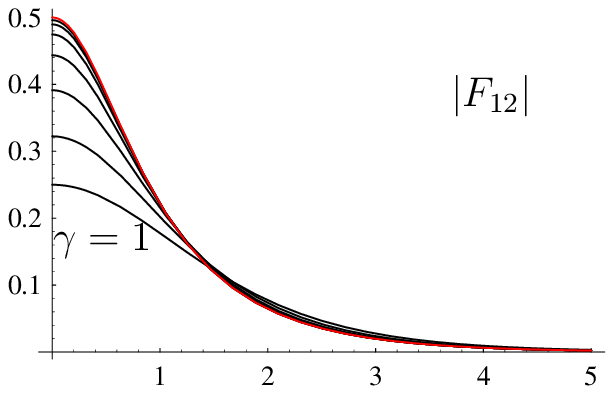}
\end{tabular}
\caption{\sf $h_{1,2}$ (solid, broken lines) and $B_{e,g}$. The left panels ($m_e = 1$) for
$\log \gamma = 0,1,2,3,4,5$ and $
\infty$. 
The right panels ($m_g = 1$) for 
$\log\gamma = 0,-0.5,-1,\cdots,-3$ and $-\infty$. 
}
\label{fig:profile}
\end{center}
\end{figure}
Numerical solutions  for $N=2$ 
for a wide range of $\gamma$ (including $\gamma=0,\infty$) 
are shown in Fig.~\ref{fig:profile}.
Winding field $h_1$ is not sensitive on $\gamma$ while unwound field $h_2$ is.
As $m_g$ being sent to $\infty$ ($\gamma \to \infty$), 
the non-Abelian flux $F_{12}$ becomes very sharp and finally gets to singular.
Interestingly, the Abelian flux $f_{12}$ is kept finite there.
In a region $\gamma < 1$ ($m_e > m_g$), on the other hand,
the Abelian flux is a bit smaller than the non-Abelian tube.
Surprisingly, the fluxes remain finite even in $m_e \to \infty$ 
limit.

\noindent
{\it 3.\ 
Asymptotic width}.
Let us investigate the vortex solution by expanding 
(\ref{eq:mm1a}) and (\ref{eq:mm2a}) in region 
$r \gg \max\{m_e^{-1},m_g^{-1}\}$ where 
$|\psi_e|,|\psi_g| \ll 1$. 
We keep only the lowest-order term in $\psi_e$ 
while keeping terms up to next to leading order in 
$\psi_g$ 
in Eq.(\ref{eq:mm1a}):
\beq
(\triangle  - m_e^2) \psi_e + \frac{m_e^2 \psi_g^2}{2(N-1)}  = 0,\ ~~
(\triangle  - m_g^2) \psi_g = 0.
\label{eq:asympt}
\eeq
The solution is given by 
the second modified Bessel function $K_0(r)$, 
and approximated as 
\beq
\psi_e \simeq
\left\{
\begin{array}{c}
c_e \sqrt{\frac{\pi}{2m_er}} e^{-m_er}, \\
\frac{\pi c_g^2}{4(N-1)(1 - 4\gamma^2)} \frac{e^{-2 m_g r}}{m_g r}, 
\end{array}
\right. \
\psi_g \simeq c_g\sqrt{\frac{\pi}{2m_g r}} e^{-m_g r},\ 
\label{eq:asym_sol}
\eeq
with $c_{e,g}$ being dimensionless constants, see Table \ref{tab:values}. 
\begin{table}[ht]
\begin{center}
\begin{tabular}{c|ccccc}
$\gamma$ & $c_e$ & $c_g$ & $b_{e}$ & $b_{g}$ & $a_\gamma$ \\
\hline
$0$ & --& 1.1363 
& 0 & 0.75905 & $\sqrt  2$ \\
$0.25$ & --& 1.1853
& 0.31719 & 0.73163 & 1.31688\\
$0.5$ & --& 1.3090
& 0.47907 & 0.68393 & 1.18361\\
$0.75$ & 2.196
& 1.4852
& 0.55921 & 0.64006 & 1.07932\\
$1$ & 1.70786 
& 1.70786 
& 0.60329 & 0.60329 & 1 \\
$1.5$ & 1.4715
& 2.3031
& 0.64726 & 0.54697 & 0.88820 \\
$2$ & 1.4037
& 3.15
& 0.66773 & 0.50604 & 0.81226\\
$2.5$ & 1.3746
& 4.32
& 0.67897 & 0.47469 & 0.75640\\
$3$ & 1.3594
& 6.0
& 0.68584 & 0.44969 & 0.71301 \\
$\infty$ & 1.3267
&-- & 0.70653 
& 0 & 0
\end{tabular}
\caption{\sf Numerical data for $k=1$ $U(2)$ vortex. 
}
\label{tab:values}
\end{center}
\end{table}
\vspace*{-.6cm}
The asymptotic behavior of $\psi_e$ changes at $\gamma = 1/2$ (upper for $\gamma \ge 1/2$ and lower $\gamma \le 1/2$). 
Similar phenomenon was observed for the non-BPS ANO 
vortex \cite{Plohr:1981cy,Perivolaropoulos:1993uj}.
The origin of $\psi_e(\psi_g)$ is (non-)Abelian fields 
$\phi_{\rm A}(\phi_{\rm N})$ with mass $m_e(m_g)$, 
and the $\gamma = 1/2$ threshold 
can be interpreted as 
follows.
The expansion of the Lagrangian with respect 
to small 
$\phi_{\rm A,N}$ 
contains the triple couplings $\phi_{\rm A} \phi_{\rm N}^2$. 
For $m_e\le 2 m_g$, asymptotics for $\phi_{\rm A,N}$ are 
given by $K_0(m_{e,g} r)$ as the two-dimensional Green's function.
When $m_e > 2 m_g$, 
the particles $\phi_{\rm A}$ decay into 
two particles $\phi_{\rm N}^2$  
through these couplings, and thus,
$\phi_{\rm A}$ exhibits the asymptotic behavior $e^{-2m_g r}$ 
below $\gamma = 1/2$ like Eq.~(\ref{eq:asym_sol}). 
On the contrary, even 
for 
$\gamma>2$, $\phi_{\rm N}$ does not behave as 
$e^{-2m_e r}$ since there is no 
triple coupling $\phi_{\rm N}\phi_{\rm A}^2$ due to 
the traceless condition for $\phi_{\rm N}$.

Let us define {\it asymptotic width of the 
vortex 
} by
an inverse of the decay constant in Eq.(\ref{eq:asym_sol}):
\beq
L_e = \left\{
\begin{array}{cc}
2/m_e & \text{for}\ 
\gamma \ge 1/2,\\
2/(2m_g) & \text{for}\ 
\gamma < 1/2,
\end{array}\right.\quad L_g = 2/m_g.
\label{eq:asym_size}
\eeq
Here the factor 2 is put in the numerator 
to match with another definition in Eq.(\ref{eq:core_width}). 
The asymptotic width of Abelian 
vortex 
is 
bigger than the non-Abelian one when $\gamma > 1$
and vice versa for $1/2 \le \gamma < 1$. 
For $\gamma=1$,
the two widths are the same.
The case $\gamma < 1/2$ indicates a significant modification, 
where the Abelian flux tube is supported by the non-Abelian flux tube.
This answers the question why the Abelian 
vortex 
does not 
collapse even in the $m_e\to\infty$ limit. 
When $\gamma \gg 1$, the thin non-Abelian flux hidden by 
fat Abelian flux cannot be 
correctly measured by $L_{g}$. 
We now turn to another definition of vortex width 
which reflect the size of the vortex core more faithfully.

\begin{figure}[ht]
\begin{center}
\begin{tabular}{c}
\includegraphics[height=6cm]{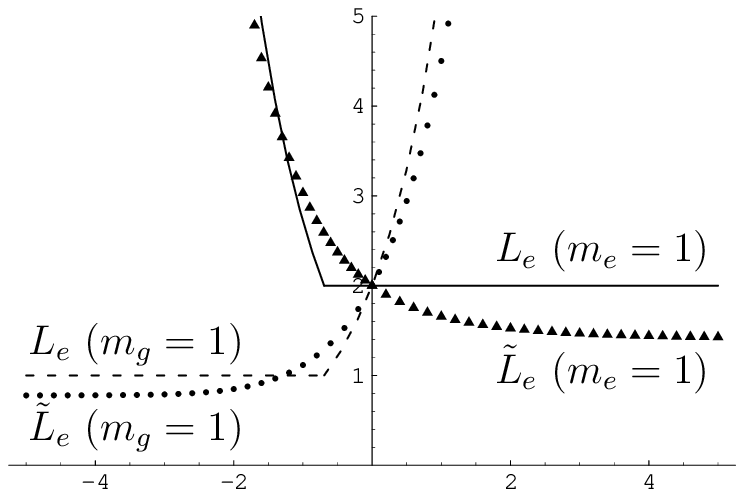} \\
\includegraphics[height=6cm]{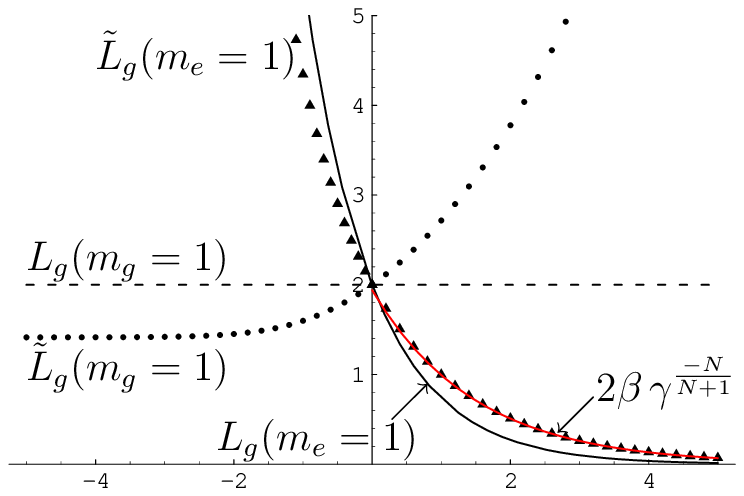}
\end{tabular}
\caption{\sf The core and asymptotic widths v.s $\log\gamma$.}
\label{fig:h2}
\end{center}
\end{figure}

\noindent
{\it 4. Core widths}.
Let us 
consider a region near the 
vortex core. We expand fields by
\beq
\psi_e \!\approx\! - \frac{2}{N}\! \log b_{e}m_e r ,\ ~~~
\psi_g \!\approx\! - 2\frac{N\!-\!1}{N}\! \log b_{g}m_g r.
\label{eq:est_origin}
\eeq
Dimensionless constants $b_{e,g}$ are related to $h_2(r=0)$ 
by 
\beq
a_\gamma \equiv h_2(0) = (b_{e}/\gamma b_{g})^{1/N}.
\eeq
See Table \ref{tab:values}.
These are important since they are related to the 
maximum values of the magnetic fluxes at $r=0$ 
\beq
B_e 
= -\frac{m_e^2}{2}\left( 1 - \frac{N-1}{N}a_\gamma^2 \right),\quad
B_g 
= - \frac{m_g^2}{2} \frac{N-1}{N} a_\gamma^2.
\label{eq:flux_origin}
\eeq
Widths of the magnetic fluxes 
can be 
estimated 
by using a step function $\Theta(x)$ as 
$F_{12} = B_g \Theta(\tilde L_g-r) T$ and 
$f_{12} = B_e\Theta(\tilde L_e-r)$
keeping amounts of the fluxes as
$|B_e| \times \pi \tilde L_e^2 = 2 \pi/N$ and
$|B_g| \times \pi \tilde L_g^2 = 2 (N-1) \pi/N$:
\beq
\tilde L_e = \frac{2}{m_e \sqrt{N-(N-1)a_\gamma^2}},~\quad
\tilde L_g = \frac{2}{m_g a_\gamma},
\label{eq:core_width}
\eeq
We call $\tilde L_e$ and $\tilde L_g$ as 
the {\it 
core widths} of the vortex. 
In the case of $\gamma = 1$, $\tilde L_e$ and $\tilde L_g$ 
coincide because of $a_{\gamma=1}=h_2 = 1$.
In Fig.~\ref{fig:h2}, we show the 
core 
widths  
numerically 
in the case of $N=2$, which are analytically reinforced 
as we will discuss.
We again observe that the Abelian core does not collapse even when $m_e \gg 1$ ($\gamma \ll 1$).

Mass dependence of the
core widths coincides with  one of the asymptotic widths 
$L_{e,g}$ 
given in Eq.~(\ref{eq:asym_size}),
except for $\tilde L_g (\gamma>1)$,
see Fig.~\ref{fig:h2}.
The asymptotic width $L_g$ is independent of $m_e$ 
whereas the core width $\tilde L_g$ depends on $m_e$. 
This is because $\tilde L_{e,g}$ more faithfully 
reflects the multilayer structure in the large intermediate 
region of $r$ for the strong coupling regime ($\gamma \gg 1$), 
to which we now turn. 

\noindent
{\it 5.\ Two strong coupling limits.}
Here we study two limits: i) $m_g \to \infty$ 
with $m_e$ fixed, 
and ii) $m_e \to \infty$ with $m_g$ fixed. 
In the former limit 
($\gamma \to \infty$ 
with $m_e$ fixed), 
all the fields with mass 
$m_g$ become infinitely heavy and  integrated out from
the theory. As a result, the original $U(N)$ gauge theory becomes the Abelian theory with 
one $SU(N)$ singlet field $B\equiv\det H$. 
Note that the target space is $\mathbb{C}/\mathbb{Z}_N$
because $U(1)$ charge of $B$ is $N$. 
Eq.\,(\ref{eq:mm2a}) is solved by $\psi_{g,\infty} = 0$
while $\psi_{e,\infty}$ is determined by Eq.\,(\ref{eq:mm1a})
\beq
\! \!
\frac{\triangle \psi_{e,\infty}}{m_e^2} = 1 - e^{-\psi_{e,\infty}}, 
~~\ \psi_{e,\infty} \underset{r \rightarrow 0}{\longrightarrow} 
- \frac{2}{N} \log m_e b_{e,\infty} r
\eeq
where suffix $\infty$ denotes $\gamma = \infty$: 
$\psi_{g,\infty}\equiv \psi_g|_{\gamma\to \infty}$. 
The boundary condition tells that vorticity is fractional 
$k=1/N$. 
This way the non-Abelian flux tube collapses and
the $U(N)$ non-Abelian vortex reduces to 
the $1/N$ fractional Abelian vortex. 
This solution helps us 
to understand the non-Abelian 
vortex for $\gamma \gg 1$ better. 
Since
$e^{-\psi_e} \approx (m_eb_{e,\infty}r)^{2/N}$ ($r \ll 1/m_e$) 
for $\gamma\gg 1$, 
$\psi_g$ for $r\ll 1/m_e$ is 
well approximated by a solution of the following
\begin{eqnarray}
 \frac{\triangle  \psi_g}{\tilde m_g^2}  
+ \frac{N-1}{N} (\tilde m_g\, r)^{\frac2N}
\left( e^{-\psi_g} - e^{\frac{\psi_g}{N-1}}\right) = 0,
\end{eqnarray}
where the parameter 
$\tilde m_g \equiv m_g (b_{e,\infty}\gamma^{-1} )^{\frac1{N+1}}$ has 
a mass dimension.
Therefore $\psi_g$ has asymptotic behavior in the middle region 
$1/\tilde m_g \ll r \ll 1/m_e $ 
\begin{eqnarray}
 \psi_g\approx \tilde c_g \, 
K_0\left(\frac{N}{N+1}(\tilde m_g r)^{\frac{N+1}N}\right)\ll 1,
\label{eq:middle}
\end{eqnarray}
and $\psi_g \approx - \frac{2(N-1)}{N}\log(\tilde b_g\, \tilde m_g  r)$ 
for $r\ll 1/\tilde m_g$. 
Here $\tilde b_g,\,\tilde c_g$ are determined numerically and independent of $\gamma$, for instance, 
$\tilde b_g=0.74672,\,\tilde c_g=0.63662$ for $N=2$.
Comparing this with Eq.\,(\ref{eq:est_origin}), we find 
$b_{g} \approx  \tilde b_g \, [b_{e,\infty}\gamma^{-1}]^{\frac1{N+1}}$
and $a_\gamma\approx \tilde b_g^{-\frac1N} \,
[b_{e,\infty}\gamma^{-1}]^{\frac1{N+1}}$ for $\gamma\gg 1$.

In the second limit 
($\gamma \to 0$ with $m_g$ fixed), 
all the fields with the mass 
$m_e$ are integrated out. 
The model reduces to 
a $\mathbb{C}P^{N^2-1}$ model 
with $SU(N)$ isometry [in $SU(N^2-1)$] gauged.
Eq.~(\ref{eq:mm1a}) is solved by
$
e^{\psi_{e,0}} = (
e^{-\psi_{g,0}} + (N-1)e^{\frac{\psi_{g,0}}{N-1}}
) / N
$ while $\psi_{g,0}$ is determined by
\beq
\triangle  \psi_{g,0} = m_g^2 \frac{ (N-1) \left(1 - e^{-\frac{N}{N-1}\psi_{g,0}} \right) }{
(N-1) + e^{-\frac{N}{N-1}\psi_{g,0}}}, 
\eeq
where the suffix $0$ denotes $\gamma\to 0$: 
$\psi_{g,0}\equiv\psi_g|_{\gamma\to 0}$. 
This is a {\it new} $\sigma$-model lump with the 
non-Abelian flux accompanied with 
the internal orientation $\mathbb{C}P^{N-1}$. 
Again we can make use of this solution to understand the 
non-Abelian vortex 
for 
$\gamma \ll 1$.
Let us define 
$\alpha^2 \equiv 
\frac{\triangle \psi_{g,0}(0)}{\triangle \psi_{e,0}(0)} 
= \frac{B_g}{B_e}\big|_{\gamma\to0}$
which 
turns out to be finite 
$\alpha^2 =(N-1)/(1+4 b_g^2|_{\gamma\to 0})$.  
Since $\triangle \psi_{g,0}(0) = m_g^2$ and
$\triangle \psi_{e,0} (0) = 
\lim_{\gamma \to0} m_g^2\gamma^{-2} (1- (N-1) a_\gamma^2 / N )$,
we find $a_\gamma = a_0\left(1 - \gamma^2/(2\alpha^2) +
\cdots\right)$, $a_0= \sqrt{N/(N-1)}$
for $\gamma \ll 1$.

\noindent
{\it 6.\ Summary and Discussion.}
We have proposed two length scales for fluxes of non-Abelian vortices:
asymptotic widths $L_{e,g}$ in Eq.~(\ref{eq:asym_size})
and core widths $\tilde L_{e,g}$ in Eq.~(\ref{eq:core_width}).
By using the asymptotics of $a_\gamma$ obtained above, the core width
is summarized as
\beq
\left\{\tilde{L}_e, \tilde{L}_g \right\} \simeq \left\{
\begin{array}{cl}
\left\{ \frac{2\alpha}{m_g  \sqrt{N}}
 \, ,\ \frac2{m_g}\sqrt{\frac{N-1}{N}}\right\}
&(\gamma \ll 1),\\
\left\{ \frac{2}{m_e\sqrt{N}} 
\ ,\ \frac{2\beta}{m_g}\left(\frac{m_g}{m_e}\right)^{\frac{1}{N+1}} 
\right\} & (\gamma \gg 1),
\end{array}
\right. \label{eq:results}
\eeq
where $\alpha$ and $\beta$ depend only on $N$ and are 
determined numerically,
for instance $\alpha=0.55010, \beta=0.97022$ for $N=2$.
The core and asymptotic widths have the same mass dependence except for
$\tilde L_g$ and $L_g$ for $\gamma \gg 1$. 
Interestingly, the Abelian flux does not collapse even when $m_e \to \infty$ ($\gamma \ll 1$).
For  $\gamma \gg 1$, the thin non-Abelian flux
is hidden by fat Abelian flux, so that the true width cannot be captured by
the asymptotics at $r \gg m_e^{-1}$.
Instead we should use improved approximation given in Eq.~(\ref{eq:middle}) to measure 
the non-Abelian flux. Indeed, the decay constant in Eq.~(\ref{eq:middle}) is $\tilde m_g^{-1}$ whose mass dependence
is the same as one of $\tilde L_g$ for $\gamma \gg 1$.

In the limit  $m_g\to 0$, the original $U(N)$ gauge theory 
reduces to $U(1)$ gauge theory coupled to $N^2$ Higgs fields.
Eq.~(\ref{eq:results}) tells us that 
the vortex is diluted and vanishes in this limit. 
This is consistent with the fact that 
there is no (smooth) vortex solution with 
a winding number $1/N$ in that $U(1)$ theory.  
The minimal vortex in the $U(1)$ theory corresponds to 
$N$ vortices in the original theory.
The dilution is expected to be avoided 
and all the fields with mass 
of the order of 
$m_g$ decouple, 
when $N$ vortices are arranged as $H=f(r) {\bf 1}_N$.
In the limit $m_e\to 0$, there is no BPS vortex solution 
since the $U(1)$ gauge field is decoupled from the Higgs fields. 
Actually, according to Eq.~(\ref{eq:results}),
one can find both of the Abelian and the non-Abelian fluxes 
are diluted again even in this limit due to the factor 
$\gamma^{\frac{1}{N+1}}$. 
Monopoles/instantons attached by vortices 
are known to exist \cite{vm}. 
The above observation implies that such configurations reduce to a 
monopole/instanton 
configurations 
in the $SU(N)$ gauge 
theory in that limit, and strongly supports the 
correspondence between the moduli spaces of them. 

It is interesting to study relation between non-BPS vortices 
and monopoles. 
It was found that monopoles do not collapse when the Higgs mass
is very large \cite{Bogomolny:1976ab}.


M.E. and K.O. would like to thank S.B.Gudnason, N.Manton, 
D.Tong and W.Vinci for useful discussions. 
This work is supported in part by Grant-in-Aid for 
Scientific Research from the Ministry of Education, 
Culture, Sports, Science and Technology, Japan No.17540237,
No.18204024 (N.S.) and No.20740141 (M.N.).
The work is also supported by the Research Fellowships of the Japan Society for
the Promotion of Science for Research Abroad (M.E. and K.O.) and
for Young Scientists (T.F. and T.N.).

\end{document}